\documentclass[preprint]{aastex}
\usepackage{color, soul}
\usepackage{graphicx}
\usepackage{subfigure}
\usepackage{rotating}

\begin{document}
\title{{\it Chandra} Detection of X-ray Emission from Ultra-compact Dwarf Galaxies and Extended Star Clusters}

\author{Meicun Hou\altaffilmark{1,2}, Zhiyuan Li\altaffilmark{1,2}}
\affil{$^{1}$ School of Astronomy and Space Science, Nanjing University, Nanjing 210046, China}
\affil{$^{2}$ Key Laboratory of Modern Astronomy and Astrophysics (Nanjing University), Ministry of Education, Nanjing 210046, China}
\email{lizy@nju.edu.cn}

\begin{abstract}
We have conducted a systematic study of X-ray emission from ultra-compact dwarf (UCD) galaxies and extended star clusters (ESCs), based on archival {\sl Chandra} observations. 
Among a sample of 511 UCDs and ESCs complied from the literature, 17 X-ray counterparts with 0.5-8 keV luminosities above $\sim$$5\times10^{36} {\rm~erg~s^{-1}}$ are identified, which are distributed in eight early-type host galaxies. 
To facilitate comparison, we also identify X-ray counterparts of 360 globular clusters (GCs) distributed in four of the eight galaxies. 
The X-ray properties of the UCDs and ESCs are found to be broadly similar to those of the GCs. 
The incidence rate of X-ray-detected UCDs and ESCs, $(3.3\pm0.8)$\%, while lower than that of the X-ray-detected GCs [($7.0\pm0.4)$\%], is substantially higher than expected from the field populations of external galaxies.  
A stacking analysis of the individually undetected UCDs/ESCs further reveals significant X-ray signals, which corresponds to an equivalent 0.5-8 keV luminosity of $\sim$$4\times10^{35} {\rm~erg~s^{-1}}$ per source.
Taken together, these provide strong evidence that the X-ray emission from UCDs and ESCs is dominated by low-mass X-ray binaries having formed from stellar dynamical interactions, consistent with the stellar populations in these dense systems being predominantly old.
For the most massive UCDs, there remains the possibility that a putative central massive black hole gives rise to the observed X-ray emission. 
\end{abstract}


\section{Introduction}
The advent of high-resolution optical imaging and follow-up spectroscopic surveys in the past two decades, has led to the recognition of a morphologically distinct class of stellar assemblies in and around external galaxies, the first few cases of which were found in the Fornax cluster \citep{Hilker1999,Drinkwater2000}. 
These so-called ultra-compact dwarfs (UCDs; \citealp{Phillipps2001}) manifest themselves as compact objects with typical effective radii of 10 $\lesssim r_{\rm eff} \lesssim 100$ pc and absolute V-band magnitudes of -14 $\lesssim M_{\rm V} \lesssim$ -9\,mag, just intermediate between the classical globular clusters (GCs) and dwarf elliptical galaxies.  
When spectroscopic information is available, UCDs appear to harbor a predominantly old stellar population (e.g., \citealp{Paudel2010}; Janz et al. 2016), in some cases with an extended star formation history (e.g., \citealp{Norris2015}).
Candidate UCDs are now routinely found in dense environments such as galaxy clusters (e.g., Fornax, \citealp{Gregg2009}; Virgo, \citealp{Hasegan2005}; Centaurus, \citealp{Mieske2007}; Coma, \citealp{Madrid2010}; Perseus, \citealp{Penny2014})
and galaxy groups (e.g., HCG\,22 and HCG\,90, \citealp{Da Rocha2011}),  
but also in relatively isolated galaxies such as the Sombrero (= M104; \citealp{Hau2009}), NGC\,3923 and NGC\,4546 \citep{Norris2011}. 

The nature of UCDs, however, is far less clear than their names might have indicated. 
Viable formation scenarios proposed for UCDs include: (i) they are the residual cores of tidally-stripped nucleated dwarf galaxies in dense environments \citep{Bekki2001,Drinkwater2003,Pfeffer2013}; 
(ii) they are the end-product of the aggregation of young massive star clusters formed during violent gas-rich galaxy mergers \citep{Fellhauer2002,Bruns2011}; 
and (iii) they are ultra-massive GCs, as expected by extension of the GC luminosity function \citep{Mieske2002,Mieske2012}. 

Adding to the mystery of UCDs is a closely-relevant, growing class of stellar systems, namely, extended stellar clusters (ESCs). With visual luminosities largely overlapping the typical range of GCs  (-9 $\lesssim M_{\rm V} \lesssim$ -5), the ESCs earn their names for 
their characterisic sizes (10 $\lesssim r_{\rm eff} \lesssim 30$ pc) that are much larger than the classical GCs.  
A number of ESCs have also been found in galactic disks (e.g., NGC\,1023 and NGC\,3384, \citealp{Blom2002}), which are not easily explained by the scenario of stripped satellites. 
Recently, Br\"{u}ns \& Kroupa (2012; hereafter BK12) compiled from the literature a catalog of 813 confirmed and candidate UCDs and ESCs. 
In reality, the known UCDs and ESCs may represent a heterogeneous family of stellar systems that the above formation scenarios conspire to form. 

Studies of UCDs and ESCs have been concentrated on optical observations and theoretical investigations. Up till now we know very little about their properties at other wavelengths, in particular the X-ray band. 
For GCs, it has long been established that they exhibit an over-abundance of low-mass X-ray binaries (LMXBs) with respect to the field (\citealp{Clark1975}; \citealp{Katz1975}), which is attributed to the efficient dynamical formation of neutron star binaries in the dense core of GCs (\citealp{Fabian1975}, \citealp{Sutantyo1975}, \citealp{Hills1976}).
X-ray surveys of extragalactic GCs, mostly accomplished by the {\it Chandra X-ray Observatory}, have revealed that on average $\sim$5\% of GCs exhibit an X-ray counterpart (presumably LMXBs) at a limiting luminosity of $L_X \sim 10^{37}{\rm~erg~s^{-1}}$, and that more massive GCs have a higher probability of hosting LMXBs \citep{Fabbiano2006}. 
If some UCDs (and ESCs) are indeed giant versions of GCs, they are naturally expected to host LMXBs. 
The first UCD reported to produce X-ray emission lies in M104 (SUCD1; \citealp{Hau2009}), which has an X-ray luminosity of $L_X \approx 10^{38}{\rm~erg~s^{-1}}$ \citep{Li2010}.
In view of the rapidly growing inventory of UCDs and ESCs, it would be interesting to examine the incidence rate of X-ray sources in these dense stellar systems, which should provide clues about their internal structure.
On the other hand, many UCDs found in galaxy clusters can well be the remnant of stripped nucleated galaxies.
The recent discovery of a candidate super-massive black hole (with a mass of $\sim$$2\times10^{7}{\rm~M_\odot}$) in the UCD of M60 (M60-UCD1; \citealp{Seth2014}), based on spatially-resolved stellar kinematics, lends strong support to this hypothesis. 
Strader et al.~(2013) identified a variable X-ray source with M60-UCD1.
X-ray emission could be an important tracer of this and other putative massive black holes embedded in UCDs. 

In this work, we conduct a systematic survey of X-ray emission from UCDs and ESCs using archival {\it Chandra} observations\footnote{During the final preparation stage of our manuscript, Pandya, Mulchaey \& Greene (2016) released preprint of a similar study. Their findings are significantly different from ours, mostly because of the different sample selection criteria employed (Section 2). When relevant, we compare our results with Pandya et al. in later Sections.}, in order to shed light on the origin and nature of these intriguing objects.
In Section 2 we describe the sample selection, data reduction and X-ray source detection. In Section 3 we present the X-ray counterparts of the UCDs and ESCs and analyze their properties in close comparison with the X-ray population found in GCs. 
Discussion and summary of our results are given in Sections 4 and 5, respectively. Quoted errors are at the 1\,$\sigma$ confidence level throughout this work unless otherwise noted.

\section{Data Preparation} \label{sec:data}

\subsection{Sample selection and data reduction} \label{subsec:X-ray data}
We adopt the BK12 catalog of UCDs and ESCs as our primary sample, which results from an exhaustive search of the literature available at the time. 
We note that only a (growing) fraction of the cataloged sources have been spectroscopically confirmed. 
Nevertheless, BK12 showed that the spectroscopically confirmed sources are statistically representative of the full sample, in particular in the $r_{\rm eff}-M_{\rm V}$ plane.
The UCDs and ESCs were collectively dubbed extended objects (EOs) in BK12 to reflect the unclear physical distinction between the two groups, a naming convention we follow in this work. 
We caution that the term ``EO" should only be treated as a technical, rather than physical, distinction from the more compact GCs. 
The 813 cataloged sources were tentatively associated with 65 host galaxies ranging from dwarfs to Milky Way-like normal galaxies, and to giant ellipticals embedded in galaxy clusters.  
We dropped sources without explicit celestial coordinates. We also neglected sources associated with Local Group galaxies or with M87, the cD galaxy of the Virgo cluster. 
Sources within the Local Group are essentially well-studied ESCs, while M87 hosts a large and still growing population of UCDs \citep{Zhang2015} that deserve a joint optical/X-ray investigation elsewhere.    
Lastly, we excluded sources located in the Perseus and Coma clusters, chiefly to avoid potential bias introduced by the relatively poor X-ray source detection sensitivity, which is expected for the large distances and substantial local background caused by the hot intra-cluster medium in these two massive galaxy clusters.
We cross-correlated the remaining sources with the {\it Chandra} archive, requiring that all sources of interest fall within 8 arcmin from the aimpoint of an ACIS-I or ACIS-S observation, to ensure good sensitivity and accurate source positioning. 
In this way we selected 449 sources from the BK12 catalog.

We then added to the BK12 sources 60 ESCs recently identified in NGC\,1023 from {\it HST} imaging (\citealp{Forbes2014}), all of which have an effective radius $\geq$ 10 pc, including an {\it $\omega$\,Cen}-like object with $r_{\rm eff}$ of 10 pc and $M_{\rm V}$ of -8.9 mag (hereafter NGC\,1023-EO1).
We have converted the $g$-band magnitudes given in Forbes et al.~(2014) into V-band magnitudes, assuming an underlying simple stellar population with age of 10 Gyr, half-solar metallicity and a Kroupa initial mass function (IMF).
We also added two recently discovered UCDs that are among the most massive UCDs ever found (M60-UCD1: \citealp{Strader2013}, \citealp{Seth2014}; M59-UCD3: \citealp{Sandoval2015}, \citealp{Liu2015}).
Our final sample consists of 511 EOs, which are distributed in 27 host galaxies. Among them, 5 galaxies had only one observation and the other 22 with multiple exposures. 
Data of the queried observations are publicly available by June 2015.

We reprocessed the {\it Chandra} data with CIAO v4.7 and the corresponding calibration files, following the standard procedure\footnote{http://cxc.harvard.edu/ciao}.
Briefly, we produced count and exposure maps in the 0.5-2 ($S$), 2-8 ({$H$}), and 0.5-8 ($B$) keV bands for each observation. 
The exposure maps were weighted by an absorbed power-law spectrum, with a photon index of 1.7 and an absorption  column density $N_{\rm H}$ = $10^{21} {\rm~cm^{-2}}$.
For galaxies with multiple observations, we calibrated their relative astrometry by matching the centroid of commonly detected point sources, using the CIAO tool {\sl reproject\_aspect}.
The count and exposure maps of individual observations were then reprojected to a common tangential point, i.e., the optical center of the putative host galaxy, to produce combined images of optimal sensitivity for source detection.
The energy-dependent effective area among the ACIS CCDs was taken into account, assuming the above incident spectrum, so that the quoted count rates throughout this work refer to ACIS-S3.

Table 1 presents basic information of the host galaxies, including position, distance, effective exposure, and the number of EOs within the {\it Chandra} field-of-view (FoV).

\subsection{X-ray source detection} \label{sec:detection}
At distances of a few Mpc and beyond, UCDs and ESCs should remain unresolved even under the superb angular resolution afforded by {\it Chandra}.
To form the basis of identifying X-ray counterpart of the UCDs/ESCs, we ran source detection in the 0.5-2 ($S$), 2-8 ($H$) and 0.5-8 ($B$) keV bands for each host galaxy, following the procedure detailed in \citet{Wang2004} and \citet{Li2010}.
The source detection algorithm has been successfully applied for detecting X-ray sources in M104 (\citealp{Li2010}), which is also included in the present study.
For each detected source, background-subtracted, exposure map-corrected count rates in the individual bands are derived from within the 90\% enclosed energy radius. 
According to the assumed incident spectrum, we have adopted a $B$-band count-rate-to-luminosity conversion factor of $9.6(d/{\rm Mpc})^2\times10^{38}{\rm~erg~s^{-1}/(\rm cts~s^{-1})}$, where $d$ is the distance to a given galaxy (Table 1).
The 0.5-8 keV detection limit for each galaxy field, ranging between a few $10^{35}$ to a few $10^{38}{\rm~erg~s^{-1}}$ (with a median of $2.0\times10^{37}{\rm~erg~s^{-1}}$), is given in Table 1.
For the galaxies with multiple observations, we repeated the above procedure in individual exposures to probe long-term source variability.

\section{Analysis and results} \label{sec:analysis}
\subsection{Identifying X-ray counterparts}
We search for X-ray counterpart of the UCDs and ESCs from the list of detected sources in each galaxy field, by adopting a matching radius of $2^{\prime\prime}$, which corresponds to a physical scale of $\sim$156 pc at a distance of 16.1 Mpc, the median of our sample galaxies. 
This choice allows for position uncertainty of sources found at relatively large off-axis angles\footnote{According to the extensive simulations of Kim et al.~(2007) for the ChaMP X-ray Point Source Catalog, sources with net counts $\lesssim$ 50 and off-axis angles of $3^\prime$-$8^\prime$ have 95\% positional uncertainties of 0\farcs5-1\farcs8.} and the (typically unknown) relative astrometry between the X-ray and optical observations.  
A total of 17 X-ray counterparts are thus identified, including 1 in NGC\,1023, 1 in NGC\,1399, 6 in NGC\,4365, 1 in M89, 1 in M59, 1 in M60, 2 in M104 (Sombrero), and 4 in NGC\,5128 (Cen A). 
Notably, all eight host galaxies are early-type galaxies, although the {\it Chandra} pointed observations might have been biased against late-type galaxies.  
To assess the probability of random matches, we artificially shift the positions of all detected sources in each field by $\pm10^{\prime\prime}$ in RA and DEC and average the number of coincident matches in the four directions. 
We find essentially zero random matches in all the fields except for NGC\,4365, in which the above exercise results in 3.25 random matches.   
This can be understood, since NGC\,4365 has both a large population (216) of EOs and a large number (369) of detected X-ray sources. 
By reducing the matching radius to $1^{\prime\prime}$, we find 11 X-ray counterparts, indicating that some of the eliminated matches could indeed be interlopers -- in particular, 4 eliminated cases are in NGC\,4365.
Nevertheless, we consider all 17 sources as true X-ray counterparts and provide their basic properties in Table 2, in which we also quote the BK12 catalog for their V-band absolute magnitude and effective radius. 
From the literature we find that at least 7 of the 17 sources (NGC\,1399-EO12, M104-EO01, M59-UCD3, M60-UCD1, NGC\,5128-EO03, NGC\,5128-EO05, NGC\,5128-EO01) are confirmed UCDs/ESCs,
for which we quote the spectroscopically-derived metallicity in Table 2; for the remaining sources, we have estimated the metallicity from their optical color, assuming a simple stellar population with age of 10 Gyr and a Kroupa IMF. 
Roughly an equal number of X-ray counterparts are found in the metal-rich ([M/H] $\geq$ 0.3) and metal-poor ([M/H] $<$ 0.3) UCDs/ESCs.
Figures 1 and 2 show the {\it Chandra} 0.5-8 keV images of the eight fields, with the positions of the identified X-ray sources marked.

Among the eight host galaxies, some have a well-documented population of GCs, which are valuable for a direct comparison with the UCDs/ESCs in the  observed X-ray properties.
For this purpose, we adopt the published GC catalogs of NGC\,1399 \citep{Blakeslee2012}, NGC\,4365 \citep{Blom2012}, M104 \citep{Spitler2006} and NGC\,5128 \citep{Harris2012}. 
Again by adopting a matching radius of $2^{\prime\prime}$, we find 101, 184, 65 and 10 X-ray counterparts for the GCs in NGC\,1399, NGC\,4365, M104\footnote{In \citet{Li2010}, we identified X-ray counterpart for 41 GCs, but there a more restrictive matching radius of 0\farcs5 was adopted.} and NGC\,5128, respectively (marked by green circles in Figures 1 and 2). 
The number of GCs within the individual FoV (5140 in total) and the number of their identified X-ray counterparts are listed in Table 1.
The total incidence rate of X-ray sources in GCs (hereafter XGCs) is $(7.0\pm0.4)$\%. 
This is to be contrasted with the $(3.3\pm0.8)$\% incidence rate of the X-ray-emitting UCDs/ESCs (hereafter collectively called XEOs) found in all {\it Chandra} fields (or $[3.1\pm0.8]$\%, if only sources in the BK12 catalog are taken into account).
We caution that our parent sample of UCDs and ESCs, resulted from literature compilation, is likely more heterogeneous than the GC sample, in terms of completeness.  

A comparison between our identifications and the X-ray identifications by Pandya et al.~(2016) is warranted. 
From their own literature compilation of UCD candidates, which are distributed primarily in galaxy clusters, Pandya et al. (2016) identified 21 X-ray counterparts by adopting a matching radius of $1.5^{\prime\prime}$,
among which six sources (NGC\,1399-EO12, M104-EO01, M60-UCD1, NGC\,5128-EO02, NGC\,5128-EO05 and NGC\,5128-EO01) are in common with our XEOs. 
The remaining 15 sources are not included in our primary sample. A close examination indicates that some of these sources, considered UCDs by Pandya et al.~(2016), have an effective radius of  3-8 pc (i.e., more typical of GCs), and thus would not have appeared in the BK12 catalog. 
On the other hand, the 11 XEOs that are identified by us but not included in Pandya et al. (2016) are essentially sources not in their parent sample.

\subsection{Global X-ray properties}

In Figure~\ref{fig:hardness}, we show the 0.5-8 keV intrinsic luminosity ($L_{\rm X}$) against hardness ratio of the 17 XEOs.
The hardness ratio, defined as $(H-S)/(H+S)$ and listed in Table 2, is calculated from the observed counts in the $S$ (0.5-2 keV) and $H$ (2-8 keV) bands, using a Bayesian approach \citep{Park2006}. 
For comparison, we also plot in Figure~\ref{fig:hardness} the 360 XGCs from four host galaxies (Section 3.1; Table 2), which are presumably LMXBs.  
None of the XEOs exhibits $L_{\rm X} > 10^{39}{\rm~erg~s^{-1}}$, i.e., the regime of ultra-luminous X-ray sources, where black hole binary systems may be relevant.
The majority of XGCs also fall short of this threshold; only four XGCs in NGC\,1399 have $L_{\rm X} > 10^{39}{\rm~erg~s^{-1}}$.
The XEOs and XGCs are also similar in the distribution of their hardness ratios. 
Notably, the ultra-massive M59-UCD3 is the softest among all XEOs. This source is detected in the $S$-band but not in the $H$-band, thus having a hardness ratio of -1. 
Visual inspection of the {\it Chandra} image indicates that all 7 photons from M59-UCD3 have an energy below 1.8 keV.  

Several XEOs have sufficient net counts for spectral analysis. For such sources, we extract their spectra from a 2$^{\prime\prime}$-radius circle, and the corresponding background spectra from a concentric ring with inner-to-outer radii of $3^{\prime\prime}$-$5^{\prime\prime}$. Spectra extracted from multiple exposures of the same source are co-added. 
All the spectra appear virtually featureless, and thus we fit them with an absorbed power-law model, requiring that the equivalent hydrogen column density is no less than the Galactic foreground value (Kalberla et al. 2005).
We obtain meaningful constraints on the photon-index for M104-EO01 (i.e., SUCD1), M60-UCD1, NGC\,1399-EO12, NGC\,4365-EO117, NGC\,5128-EO1 and NGC\,5128-EO5, with best-fit values of $1.22^{+0.20}_{-0.20}$, $1.86^{+0.18}_{-0.18}$, $1.90^{+0.18}_{-0.18}$, $1.21^{+0.11}_{-0.11}$, $1.58^{+0.05}_{-0.05}$ and $1.78^{+0.11}_{-0.11}$, respectively.
These values are consistent with the typical range of LMXBs.

We also examine the long-term flux variability of XEOs, based on the source count rates measured from individual observations. 
Following \citet{Li2010}, we define source variability $\tilde{V} = F_{\rm h}/F_{\rm l}$, where $F_{\rm h}$ is the highest count rate among individual detections and $F_{\rm l}$ the statistical upper limit of the lowest detected count rate. 
The values of $\tilde{V}$ are listed in Table 2. A value of $\tilde{V}=1.0$ is given for the two XEOs with only one observation, while NGC\,4365-EO039 has an ill-defined $\tilde{V}$ because it is only detected in the combined image.
The remaining XEOs show moderate ($\tilde{V} < 10$) to strong ($\tilde{V} \geq 10$) variability, with the strongest variability found in NGC\,5128-EO01 ($\tilde{V} \approx 50$).
We note that one XEO, NGC\,4365-EO006, has been identified in only one of the six observations available for NGC\,4365; its flux was apparently too low to be detected in the other five exposures as well as in the combined image. 
The prevalence of flux variability in the XEOs, just as in the XGCs (\citealp{Li2010}), suggest that the bulk of the detected X-ray emission arises from a single source rather than superposition of multiple sources. 

Figure~\ref{fig:reffMV} shows the effective radius ($r_{\rm eff}$) versus absolute V-band magnitude ($M_{\rm V}$) for the 511 UCDs and ESCs studied in this work. 
By definition, all UCDs and ESCs considered here have a size lower limit of 10 pc. 
The apparent paucity of objects around $M_{\rm V} \approx -9$ might have arisen from selection effect due to the heterogeneous nature of the BK12 catalog, but otherwise can be viewed as a technical division between UCDs and the less luminous ESCs (see discussion in Forbes et al.~2013).
BK12 noticed that the majority of EOs in late-type galaxies have $M_{\rm V} > -9$.
We highlight the XEOs with red diamonds in Figure~\ref{fig:reffMV}, which distribute rather evenly across the entire range of $M_{\rm V}$. 
Ten of the 17 XEOs have $M_{\rm V} < -9$, whereas six of the 9 most luminous EOs (with $M_{\rm V} < -12$) remain undetected in X-rays. 

For comparison, we show in Figure~\ref{fig:reffMV} the GCs of M104, among which those with an X-ray counterpart are further highlighted by blue squares.
GCs in the other three galaxies are not shown due to the lack of available V-band magnitudes, but we expect that the M104 GCs are representative. 
We note that the great majority (94\%) of the GCs have $r_{\rm eff} <$ 4 pc.  
The median V-band magnitude of the GCs is -8.20 mag, and 52 out of the 65 XGCs (i.e., 80\%) are found in the brighter half.  
This clearly indicates that more massive GCs are more likely to host an LMXB, a familiar trend already noted by many previous work \citep{Fabbiano2006}.


\subsection{Stacking undetected UCDs and ESCs}
The great majority of EOs in our sample remain individually undetected in X-rays, partly owing to the relatively high limiting luminosity for most galaxies, which, at face value, is a good fraction of the Eddington luminosity of neutron star binaries. 
We employ a stacking analysis for the undetected objects to shed light on their average X-ray properties.
To do so, we collect the 0.5-8 keV counts registered within a $5^{\prime\prime}\times5^{\prime\prime}$ box around each source of interest.
These can be a subset of our total sample, e.g., EOs in a single galaxy.
An EO is excluded if it is located within $8^{\prime\prime}$ from an already detected X-ray source, to minimize contamination from PSF-scattered photons.
Next, we measure signals from the stacked count image. After several tests we choose to accumulate the total on-source counts within a $2^{\prime\prime}$-radius circle,
and estimate the background counts within a ring with inner-to-outer radii of $3^{\prime\prime}$-$5^{\prime\prime}$, after scaling the enclosed area.
The signal-to-noise radio (S/N) is calculated accordingly. We have also measured the cumulative exposure time in a similar fashion.

We examine the stacked signals of individual galaxies with at least 10 EOs (Table 1).
However, none of these galaxies alone gives a S/N $\geq$ 3. 
We further stack all undetected EOs from NGC\,1399, NGC\,4365, M104 and NGC\,5128 to enhance the S/N, which results in an average count rate of $(1.6\pm0.2)\times10^{-6}{\rm~cts~s^{-1}}$. 
Defining a V-band luminosity-weighted mean distance, $\bar{d} = [\sum_i (L_{\rm V,i}/ d^2_i)/\sum_i (L_{\rm V,i})]^{-\frac{1}{2}}$, where $L_{\rm V,i}$ and $d_i$ are the V-band luminosity and distance of the $i$th EO, the above count rate corresponds to an equivalent 0.5-8 keV luminosity of $(3.8\pm0.5)\times10^{35}{\rm~erg~s^{-1}}$ for a distance of $\bar{d} = 15.8$ Mpc.
Similarly, stacking all undetected GCs from the four galaxies, we obtain an average count rate of $(2.3\pm0.1)\times10^{-6}{\rm~cts~s^{-1}}$, or an equivalent 0.5-8 keV luminosity of  $(5.4\pm0.2)\times10^{35}{\rm~erg~s^{-1}}$ per GC. 
This suggests that the individually undetected EOs and GCs have on-average comparable X-ray emission. 


\section{Discussion} \label{sec:discussion}
The similarity in the X-ray properties (luminosity, spectra and variability) of the XEOs and XGCs (Section 3.2) strongly suggests that LMXBs also dominate the X-ray emission from EOs, if the presumption that the XGCs are essentially LMXBs holds. 
Pandya et al. (2016) drew a similar conclusion.
The presence of LMXBs in the dense, predominately old stellar systems of EOs has been naively expected. 
However, unlike the statistical behavior of the XGCs, the most luminous XEOs do not show a clear tendency of hosting an X-ray source (Figure~\ref{fig:reffMV}). 
The overall incidence rate of XEOs is also substantially low than that of the XGCs ($\sim$3\% vs. 7\%; Section 3.1). 
These findings can be understood as follows. 

It has long been recognized that the abundance (i.e., number per unit stellar mass) of luminous LMXBs in GCs is about two orders of magnitude higher than that of the Galactic field (\citealp{Clark1975}; \citealp{Katz1975}). 
This over-abundance is widely accepted as the result of stellar dynamical interactions in the dense core of GCs, where an isolated neutron star (NS) can be captured by a main sequence star through tidal force (\citealp{Fabian1975}), by a giant star through collision (\citealp{Sutantyo1975}), or by a primordial binary through exchange scattering (\citealp{Hills1976}). 
All these processes are governed by the so-called stellar encounter rate, $\Gamma \propto {\rho_c}^{2}r^3_c/\sigma$, where $\rho_c$ is the core stellar density, $r_c$ is the core radius and $\sigma$ the velocity dispersion.
The core stellar density of GCs can be as high as $10^4$-$10^5{\rm~M_\odot}{\rm~pc^{-3}}$, compared to the typical mass density of 0.1-1${\rm~M_\odot}{\rm~pc^{-3}}$ in the field, and it is this crucial factor that leads to the high incidence rate of LMXBs in GCs.

The UCDs and ESCs are also dense stellar systems, but are less so than the GCs. This is demonstrated by the dashed lines in Figure~\ref{fig:reffMV}, which mark equal values of the effective luminosity density, defined as $\rho_{\rm L} \equiv L_{\rm V}$/${r^3_{\rm eff}}$.
Approximately, one may take the effective luminosity density as a proxy of the stellar mass density\footnote{Most stellar encounters in GCs should occur within the core radius, thus this estimate of stellar density based on the effective radius is biased low for GCs. For the EOs this approximation could be more reliable, although current observations still lack the resolution to preclude the existence of a small dense core in most EOs.}.
Most GCs show $\rho_{\rm L} > 10^4 L_{\rm V, \odot}{\rm pc^{-3}}$, incidentally a threshold above which no UCD/ESC exists. 
Notably, all but three XGCs have $\rho_{\rm L} > 10^4 L_{\rm V, \odot}{\rm pc^{-3}}$, consistent with the scenario of LMXBs having formed from dynamical interaction in GCs.  
On the other hand, the XEOs can now be divided into two groups: those with $\rho_{\rm L} < 10^2 L_{\rm V, \odot}{\rm pc^{-3}}$ and those with $10^2 < \rho_{\rm L} < 10^4 L_{\rm V, \odot}{\rm pc^{-3}}$. 
Members of the latter group all lie at $M_{\rm V} \lesssim -9$ mag, i.e., they are practically UCDs. In particular, M59-UCD3 and M60-UCD1 belong to this group, and indeed they have the highest $\rho_{\rm L}$ of all UCDs, implying that the X-ray counterparts of these two ultra-massive UCDs are also LMXBs. 

We find that the incidence rate of the XEOs with $10^2 < \rho_{\rm L} < 10^4 L_{\rm V, \odot}{\rm pc^{-3}}$ is $(6.0\pm2.3)$\%, coming much closer to that of the XGCs.  
The likely reason is that the square dependence of stellar encounter rate on $\rho_{\rm L}$ is partially compensated by the larger size of the UCDs (a cubic dependence on ${r_{\rm eff}}$). 
In the same regard, the presence of XEOs with $\rho_{\rm L} < 10^2 L_{\rm V, \odot}{\rm pc^{-3}}$ is rather surprising, because the stellar encounter rate in these objects would be a factor of $10^2-10^4$ further lower. 
Recall that some of the 6 XEOs found in NGC\,4365, typically with $\rho_{\rm L} < 10^2 L_{\rm V, \odot}{\rm pc^{-3}}$, might be interlopers rather than true associations (Section 3.1).
To investigate this issue further, we repeat the above stacking analysis (Section 3.3) for two subgroups of the individually undetected EOs, one with $\rho_{\rm L} < 10^2 L_{\rm V, \odot}{\rm pc^{-3}}$  and the other with $\rho_{\rm L} > 10^2 L_{\rm V, \odot}{\rm pc^{-3}}$. 
Both subgroups show statistically significant signals, with average count rates of $(1.8\pm0.2)\times10^{-6}$ and $(7.4\pm0.6)\times10^{-6}$ ${\rm~cts~s^{-1}}$ per EO, and equivalent 0.5-8 keV luminosities of $(0.25\pm0.03)$ and $(2.0\pm0.2)\times10^{36}{\rm~erg~s^{-1}}$, respectively.
That the less dense subgroup has a lower average luminosity matches our anticipation, and also suggests that even the EOs of the lower stellar densities can have a sizable population of dynamically-formed LMXBs. 
Indeed, if dynamical effects are irrelevant, as is the case in the field, the expected number of (field) LMXBs is $N(>10^{37}{\rm~erg~s^{-1}}) = 14.3\pm8.4{\rm~per~10^{10}~M_\odot}$, empirically derived from the LMXB populations in the Milky Way and nearby galaxies (\citealp{Gilfanov2004}). 
This relation predicts a negligibly small number of $1.2\pm0.7$ XEOs, if we sum up the V-band light from all EOs and assume a V-band mass-to-light ratio of 3.3, appropriate for a simple stellar population with age of 10 Gyr and half-solar metallicity.  

In the above discussion we have neglected the role of stellar-mass black holes (BHs), which are usually thought to be absent in GCs due to their early segregation and subsequent mutual scattering at the core, although growing evidence now suggest that stellar-mass BHs do exist in some GCs (\citealp{Maccarone2007}; \citealp{Strader2012}). 
The EOs, in particular the massive UCDs which might be the remnant of stripped galaxies (e.g., \citealp{Bekki2001,Drinkwater2003}),  can harbor stellar-mass BHs. 
We note that the X-ray luminosities of all the XEOs are compatible with NS binaries (Section 3.2) and do not seemingly require the presence of BH binaries. 
Potentially also relevant is the stellar velocity dispersion. Compared to GCs, the larger velocity dispersion of EOs not only affects their stellar encounter rate, but also helps retain some of the otherwise escaping NSs and BHs.  A more quantitative treatment of all these affects is premature at this stage. 

The presence of LMXBs in EOs implies that the underlying stellar population is an evolved one. 
The dynamical formation timescale of an NS binary via tidal capture, following Hut \& Verbunt (1983), is $\tau \approx 0.4[(N_{\rm NS}/10{\rm~pc^{-3}})(N_*/10^3{\rm~pc^{-3}})(M_{\rm NS}+M_*)/{\rm M_\odot} (3R_*/R_\odot)(30{\rm~km~s^{-1}}/\sigma)(r_{\rm eff}/10{\rm~pc})^3]^{-1}{\rm Gyr}$, where $M_* \approx {\rm M_\odot}$ and $R_* \approx R_\odot$ are the characteristic stellar mass and radius, and $M_{\rm NS} = 1.4 {\rm~M_\odot}$ is the mass of NS. 
This can be regarded as an independent evidence of EOs being predominantly old stellar systems, consistent with existing optical spectroscopic studies.
One caveat of this conclusion is that our {\it Chandra} sample is biased against EOs found in disk, typically late-type galaxies (Section 3.1). 
Such EOs as possible descendants of recently aggregated massive star clusters (e.g., \citealp{Fellhauer2002,Bruns2011}) might not have sufficient time to form LMXBs,
although they are also unlikely to harbor high-mass X-ray binaries unless with very recent star formation. This latter case can be tested using optical observations.

While we have shown that the observed properties of the XEOs can be reasonably understood as them being LMXBs, the alternative possibility that some of the XEOs trace the X-ray emission from an embedded massive black hole (MBH) should not be easily dismissed.  M60-UCD1 has been shown to harbor a MBH of $2\times10^{7}{\rm~M_\odot}$ (Seth et al.~2014), and we find it to be an X-ray source with $L_{\rm X} \approx 1.1\times10^{38}{\rm~erg~s^{-1}}$ (Table 2). While the current X-ray data does not unambiguously relate the detected X-ray emission to the MBH (see also Pandya et al. 2016), we can infer an upper limit for its Eddington ratio, $\sim10^{-6.5}$, assuming that the X-ray band typically accounts for 10\% of the MBH's bolometric luminosity. 
Likewise, the ultra-massive M59-UCD3 is detected with a rather moderate $L_{\rm X} \approx 3.0\times10^{38}{\rm~erg~s^{-1}}$, but has an atypical soft X-ray spectrum, whose nature remains to be understood with enhanced S/N. Finally, we note that another ultra-massive UCD, M59cO (Chilingarian \& Mamon 2008), is undetected in the same {\it Chandra} data of M59-UCD3.  
Future high-sensitivity X-ray and optical observations should continue to provide important clues to the existence of MBHs as well as dynamical structure in the UCDs and ESCs.

\section{Summary} \label{sec:summary}
We have presented a systematic study of X-ray emission from UCDs and ESCs based on archival Chandra observations. The main results in this paper are as follows:

\begin{itemize}
\item A total of 17 X-ray counterparts are identified with 0.5-8 keV luminosities above $\sim$$5\times10^{36} {\rm~erg~s^{-1}}$, which are distributed in eight early-type host galaxies. In the meantime, 360 X-ray counterparts of GCs are identified in four of the eight host galaxies. The incidence rate of X-ray sources in UCDs/ESCs and GCs are $(3.3\pm0.8)$\% and $(7.0\pm0.4)$\% respectively.

\item The spectral and temporal properties of the X-ray-detected UCDs/ESCs are broadly similar to the X-ray-detected GCs, and are typical of LMXBs.


\item A stacking analysis further shows that there is on-average substantial X-ray emission from the individually non-detected UCDs and ESCs, which is again comparable to that from the individually undetected GCs.

\item The X-ray properties of UCDs/ESCs strongly suggest that they harbor a sizable population of LMXBs that have been formed from stellar dynamical interactions, consistent with the stellar populations in these dense systems being predominantly old. 

\item For the most massive UCDs, there remains the possibility that a central massive black hole produces the detected X-ray emission. Future high-resolution, high-sensitivity X-ray and optical observations of a carefully selected sample of UCDs and ESCs hold promise to solving their internal structure and dynamics.
\end{itemize}

\vspace{0.2cm}
This work is supported by the National Natural Science Foundation of China under grant 11133001.
The authors wish to thank Eric Peng and Yanmei Chen for helpful comments. M.H. is grateful to the hospitality of KIAA/PKU during her summer visit. Z.L. acknowledges support from the Recruitment Program of Global Youth Experts. 

\begin{figure*}\centering
\includegraphics[width=\textwidth,angle=0]{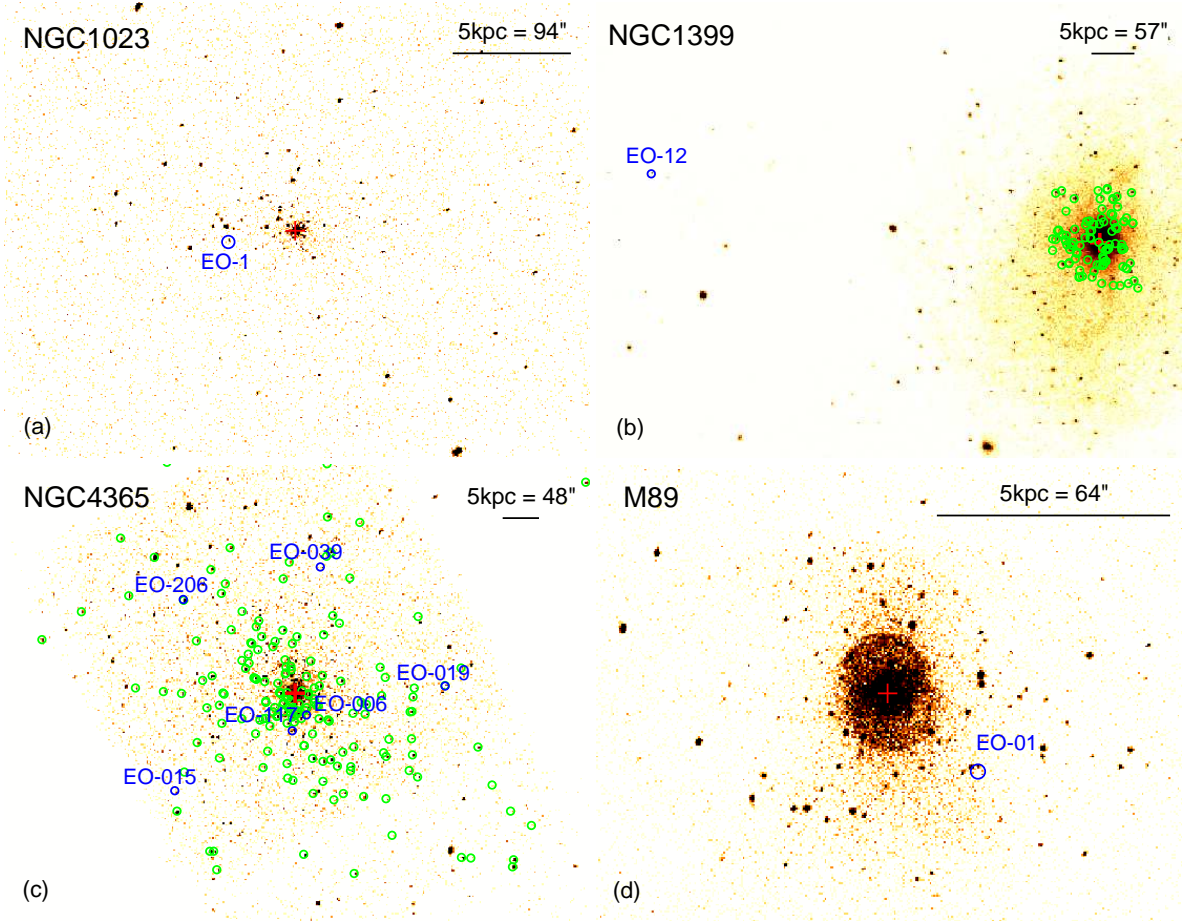}
\caption{{\it Chandra} 0.5-8 keV images of galaxies hosting X-ray-detected UCDs and ESCs: 
(a) NGC\,1023; (b) NGC\,1399; (c) NGC\,4365 and (d) M89. The UCDs and ESCs detected in X-rays are marked by blue circles. 
GCs detected in X-rays are marked by green circles. The center of each host galaxy is marked by a red cross.
}
\end{figure*}

\begin{figure*}\centering
\includegraphics[width=\textwidth,angle=0]{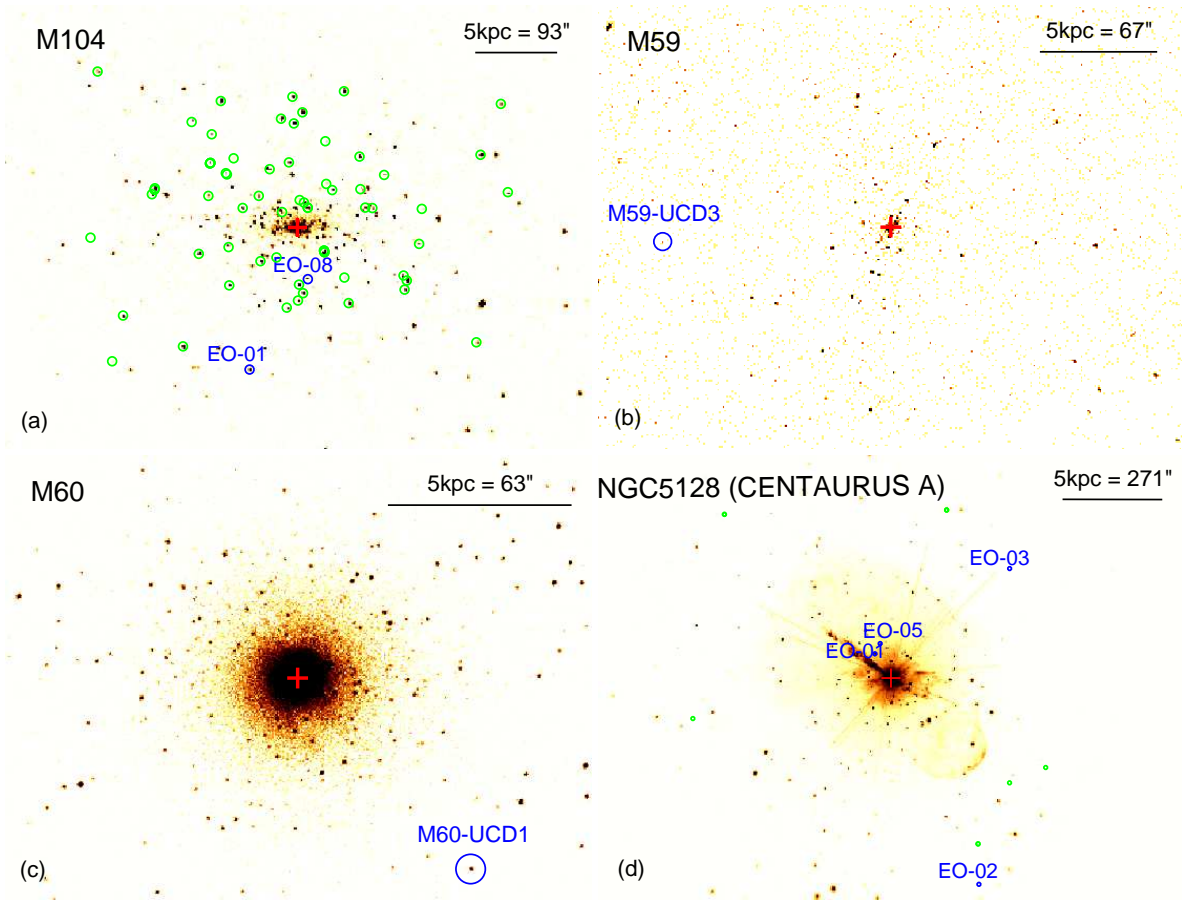}
\caption{Same as Figure 1, but for (a) M104; (b) M59; (c) M60 and (d) NGC\,5128. 
}
\end{figure*} 

\begin{figure*}\centering
\includegraphics[width=\textwidth,angle=90]{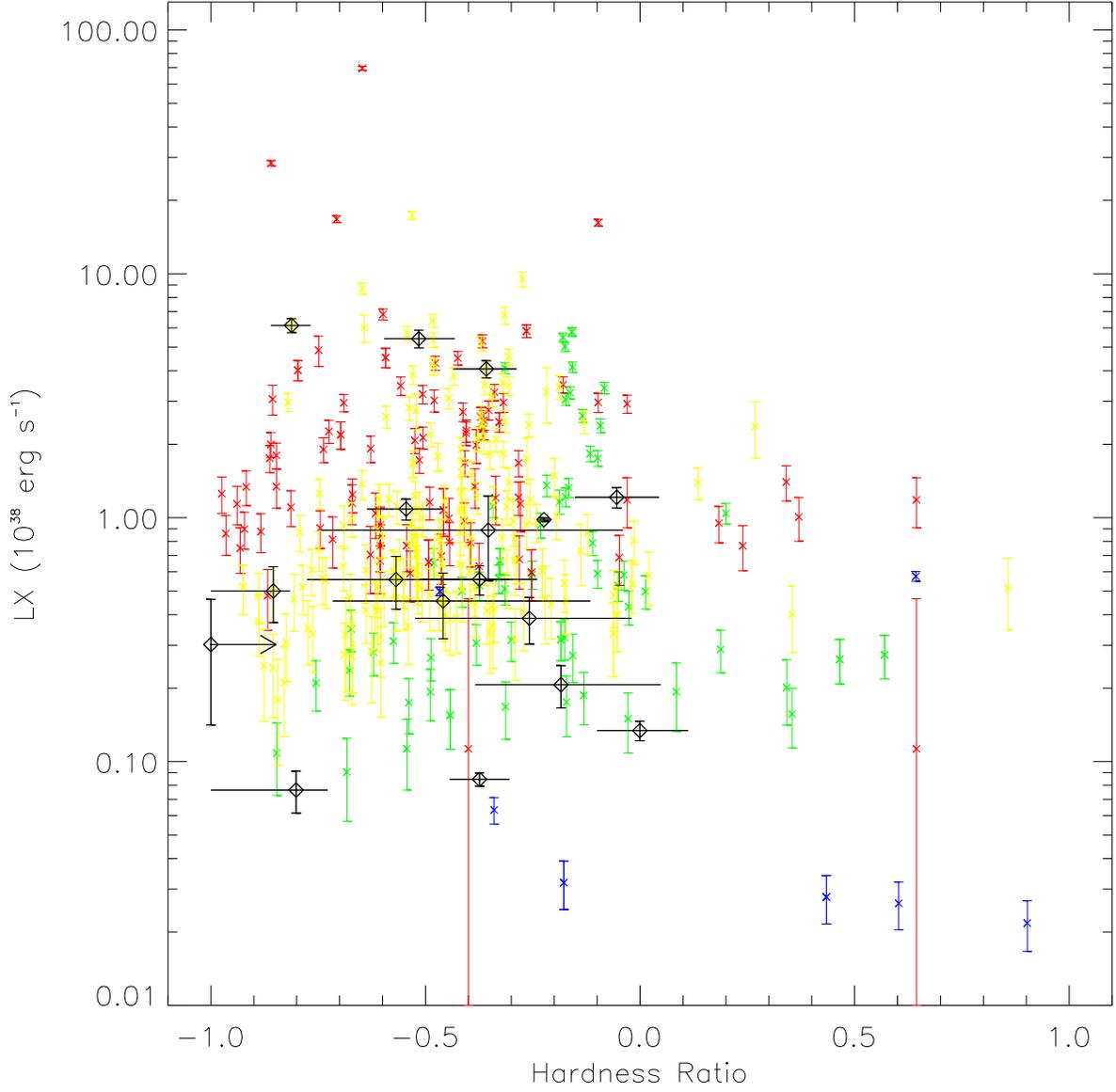}
\caption{0.5-8 keV luminosity versus hardness radio. Black diamonds represent the 17 X-ray counterparts of UCDs and ESCs. For comparison, the X-ray counterparts of GCs in NGC\,1399, NGC\,4365, M104 and NGC\,5128 are shown by red, yellow, green and blue crosses, respectively. The error bars of the GC hardness radio are neglected for clarity.}
\label{fig:hardness}
\end{figure*}

\begin{figure*}\centering
\includegraphics[width=\textwidth,angle=90]{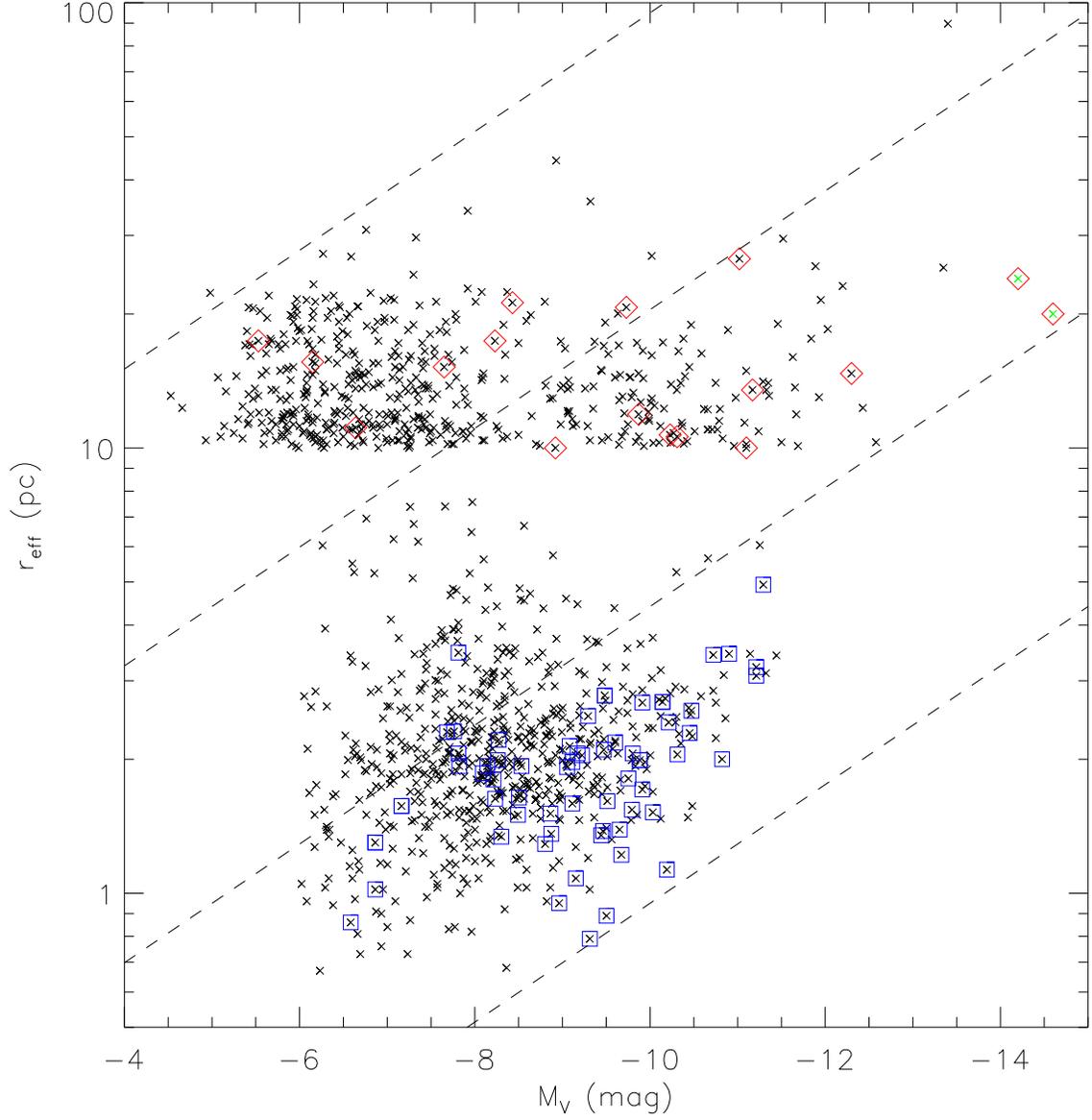}
\caption{Effective radius versus absolute V-band magnitude of EOs and GCs. All EOs, by definition, have $r_{\rm eff}$ above 10 pc, while the sources with $r_{\rm eff}$ smaller than 10 pc are GCs of M104 (Spilter et al.~2006). 
M59-UCD3 and M60-UCD1 are shown by the two green crosses. EOs (GCs) further marked by red diamonds (blue squares) have an identified X-ray counterpart. 
The multiple dashed lines indicate equal luminosity densities (defined as $L_{\rm V}$/${r^3_{\rm eff}}$) of 1, $10^2$, $10^4$, $10^{6}$ $L_{\rm V, \odot}{\rm pc^{-3}}$, from top to bottom.}
\label{fig:reffMV}
\end{figure*}



\begin{deluxetable}{cccccccccc}
\tabletypesize{\footnotesize}
\tablecaption{Basic information of EO-hosting galaxies}
\tablewidth{0pt}
\tablehead{
\colhead{Galaxy name} &
\colhead{RA} &
\colhead{DEC} &
\colhead{Dis.} &
\colhead{Exp. ($N_{\rm obs}$)} &
\colhead{$L_{\rm X, lim}$} &
\colhead{${N_{\rm EO}}$} &
\colhead{${N_{\rm XEO}}$} &
\colhead{${N_{\rm GC}}$} &
\colhead{${N_{\rm XGC}}$}\\
\colhead{(1)} &
\colhead{(2)} &
\colhead{(3)} &
\colhead{(4)} &
\colhead{(5)} &
\colhead{(6)} &
\colhead{(7)} &
\colhead{(8)} &
\colhead{(9)} &
\colhead{(10)} 
}
\startdata
NGC\,247&	11.785625&	-20.760389&	3.6&	 10.0 (2)&	5.8$\times 10^{36}$& 	2\\
NGC\,891&	35.639224&	42.349146&	10.0&	171.6 (3)&	5.2$\times 10^{36}$& 6\\
NGC\,1023&	40.1000421&  39.0632850& 11.0&  200.9 (5)& 5.5$\times 10^{36}$& 60& 1& \\
NGC\,1316&	50.673750&	-37.208056&	19.9&	20.0 (1)&	6.3$\times 10^{37}$& 45\\
NGC\,1380&	54.113750&	-34.976028&	18.3&	41.6 (1)&  4.5$\times 10^{37}$& 13\\
NGC\,1399&	54.620941&	-35.450657&	18.2&	489.7 (14)&	3.7$\times 10^{37}$& 14& 1& 401& 101\\
M81&		148.888221&	69.065295&	3.7&	 383.5 (25)&	 5.1$\times 10^{35}$& 44\\
NGC\,3115&	151.308250&	-7.718583&	9.8&	 1138.6 (11)&	 1.4$\times 10^{36}$& 5\\
NGC\,3311&	159.175000&	-27.527500&	53.7&	31.9 (1)&	7.7$\times 10^{38}$& 19\\
NGC\,3923&	177.757059&	-28.806017&	21.0&	102.1 (2)&	3.0$\times 10^{37}$& 3\\
NGC\,4278&	185.028434&	29.2807561&	16.1&	580.1 (9)&	5.8$\times 10^{36}$& 1\\
NGC\,4365&	186.117852&	7.317673&	21.4&	195.8 (6)&	2.0$\times 10^{37}$& 216& 6& 3922& 184\\
M84&		186.265597&	12.886983&	16.7&	117.2 (4)&	2.0$\times 10^{37}$& 1\\
NGC\,4382&	186.350451&	18.191487&	15.2&	49.9 (3)&	2.8$\times 10^{37}$& 4\\
NGC\,4406&	186.548928&	12.946222&	16.1&	39.8 (3)&	9.5$\times 10^{37}$& 2\\
NGC\,4449&	187.046261&	44.093630&	3.8&	 100.9 (3)&	 9.7$\times 10^{35}$& 7\\
NGC\,4472&	187.444841&	8.000476&	15.8&	462.0 (10)&	1.4$\times 10^{37}$& 1\\
M89&		188.915864&	12.5563414&	16.0&	201.4 (4)&	1.1$\times 10^{37}$& 2& 1&  & \\
M104&		189.997633&	-11.623054&	11.1&	194.2 (4)&	9.0$\times 10^{36}$& 10& 2& 659& 65\\
IC\,3652&		190.243750&	11.184556&	15.2&	5.1 (1)&	1.8$\times 10^{38}$& 1\\
M59&	190.509348&	11.647027&	15.5&	30.1 (2)&	3.8$\times 10^{37}$& 1& 1&\\
M60& 190.916564& 11.552706& 16.6& 307.9 (6)& 1.3$\times 10^{37}$& 1& 1&\\
NGC\,4660&	191.132917&	11.190306&	16.4&	5.1 (1)&	2.1$\times 10^{38}$& 1\\
NGC\,4696&	192.205208&	-41.310833&	37.6&	779.3 (15)&	1.2$\times 10^{38}$& 2\\
NGC\,5128&	201.365063&	-43.019113&	3.8&	 843.2 (24)&	 6.6$\times 10^{35}$& 16& 4& 158& 10\\
M51&		202.484200&	47.230600&	8.0&	856.6 (14)&	1.3$\times 10^{36}$& 21\\
NGC\,5846&	226.622017&	1.605625&	26.9&	149.9 (3)&	6.7$\times 10^{37}$& 13\\
\enddata
\tablecomments{(1) Name of galaxies hosting UCDs/ESCs; (2)-(3): Celestial coordinates of the galactic center (J2000); (4) Galaxy distance in units of Mpc, from the NASA/IPAC Extragalactic Database; (5) Total {\it Chandra} effective exposure, in units of ks. The number of individual observations is given in the parenthesis; 
(6) 0.5-8 keV limiting luminosity for X-ray source detection, in units of ${\rm~erg~s^{-1}}$; (7) Number of UCDs/ESCs within the FoV; (8) Number of UCDs/ESCs with X-ray counterpart, blank for zero; (9) Number of GCs within the FoV; (10) Number of GCs with X-ray counterpart. 
The GCs are from NGC\,1399 \citep{Blakeslee2012}, NGC\,4365 \citep{Blom2012}, M104 \citep{Spitler2006} and NGC\,5128 \citep{Harris2012}.
}
\end{deluxetable}

\begin{deluxetable}{cccccccccccc}
\tabletypesize{\tiny}
\tablecaption{X-ray properties of identified UCDs and ESCs}
\tablewidth{0pt}
\tablehead{
\colhead{Object Name} &
\colhead{RA} &
\colhead{DEC} &
\colhead{XRA} &
\colhead{XDEC} &
\colhead{CR} &
\colhead{L(0.5-8)} &
\colhead{$\tilde{V}$} &
\colhead{HR} &
\colhead{$r_{\rm eff}$} &
\colhead{$M_{\rm V}$} &
\colhead{[M/H]}\\
\colhead{(1)} &
\colhead{(2)} &
\colhead{(3)} &
\colhead{(4)} &
\colhead{(5)} &
\colhead{(6)} &
\colhead{(7)} &
\colhead{(8)} &
\colhead{(9)} &
\colhead{(10)} &
\colhead{(11)} &
\colhead{(12)} 
}
\startdata
NGC\,1023-EO1&	40.1191& 	39.0608& 40.1192& 	39.0610& $1.8\pm0.4$&  $2.1\pm0.41$&   2.1& $-0.18^{+0.23}_{-0.20}$  & 10.0&	-8.92& 2.0\\ 
NGC\,1399-EO12&	54.82383&	-35.42506&	54.82373& -35.42503& $17.0\pm1.4$& $54\pm4.5$& 33.0& $-0.52^{+0.08}_{-0.08}$ & 10.0&	-11.10& 0.40\\ 
NGC\,4365-EO019& 186.06212&	07.32031&	186.06209& 07.32052&$1.3\pm0.3$&  $5.6\pm1.4$&   1.4& $-0.57^{+0.26}_{-0.21}$ & 17.4&	-8.23& 0.3\\ 
NGC\,4365-EO039& 186.10850&	07.36425&	186.10835& 07.36376&$1.0\pm0.3$&  $4.6\pm1.4$&   - & $-0.46^{+0.34}_{-0.26}$ & 15.2&	-7.65& 0.15\\  
NGC\,4365-EO006& 186.11362&	07.30964& 186.11357& 07.30957&	$2.0\pm0.77$& $8.9\pm3.4$&  10.9& $-0.35^{+0.31}_{-0.39}$ &  20.7&  -9.73& 0.15\\ 
NGC\,4365-EO117& 186.11893&	07.30381&	186.11858& 07.30391& $9.2\pm0.7$&  $41\pm3.3$& 8.1& $-0.36^{+0.07}_{-0.08}$ & 11.1&	-6.64& 0.15\\ 
NGC\,4365-EO206& 186.15944&	07.35225&	186.15929& 07.35184& $13.9\pm0.9$& $61\pm4.1$& 10.9& $-0.81^{+0.04}_{-0.05}$ &17.4&-5.53&  0.3\\ 
NGC\,4365-EO015& 186.16258&	07.28169&	186.16245& 07.28119& $1.1\pm0.3$&  $5.0\pm1.3$& 1.0& $-0.85^{+0.04}_{-0.15}$ &   21.2&	-8.43& 0.02\\  
M89-EO1&	    188.90877&	12.55032&	188.90878& 12.55081& $1.6\pm0.3$&  $3.9\pm0.84$&  1.4& $-0.26^{+0.24}_{-0.27}$ & 26.6&	-11.02& 0.2\\  
M104-EO08&	189.99429&	-11.63936&	189.99474& -11.63904& $4.7\pm0.6$&  $5.6\pm0.76$& 5.0& $-0.37^{+0.13}_{-0.14}$ &  15.6&	-6.15&  0.1\\  
M104-EO01&	190.01304&	-11.66786&	190.01315& -11.66781& $10.2\pm1.0$& $12\pm1.2$&  9.3& $-0.05^{+0.10}_{-0.10}$ & 14.7&	-12.30&  0.83 \\         
M59-UCD3& 190.54605& 11.64479& 190.54615& 11.644546& $1.3\pm0.7$& $3.0\pm1.6$&  1.0& $-1.0^{+0.15}$ & 20& -14.6& 0.98\\
M60-UCD1& 190.89987& 11.53464& 190.89977& 11.534675& $4.1\pm0.4$& $11\pm1.1$& 5.6& $-0.55^{+0.08}_{-0.09}$ & 24& -14.20&  0.95\\
NGC\,5128-EO03&	201.24246&	-42.93619&	201.24250& -42.93621& $5.5\pm1.1$&  $0.76\pm0.15$& 1.8& $-0.80^{+0.07}_{-0.20}$ &   10.7& -10.23&  0.39\\
NGC\,5128-EO02&	201.27383&	-43.17519&	201.27388& -43.17508& $9.7\pm0.9$&  $1.3\pm0.12$& 4.6& $-0.00^{+0.11}_{-0.10}$ & 10.6& -10.31&  -\\
NGC\,5128-EO05&	201.37621&	-42.99300&	201.37629& -42.99297& $6.1\pm0.4$&  $0.85\pm0.05$&  9.3& $-0.37^{+0.07}_{-0.07}$&  11.9& -9.87&  0.013\\
NGC\,5128-EO01&	201.38167&	-43.00078&	201.38176& -43.00079& $70.7\pm1.2$& $9.8\pm0.17$& 49.1& $-0.22^{+0.02}_{-0.02}$ & 13.5& -11.17&  0.1\\
\enddata
\tablecomments{(1) UCDs/ESCs with X-ray counterpart. Naming convention follows BK12, except for NGC\,1023-EO1, M59-UCD3 and M60-UCD1; (2)-(3): Source position from BK12 (J2000); 
(4)-(5): Centroid position of the X-ray counterpart;
(6) 0.5-8 keV observed count rate, in units of ${10^{-4}\rm~cts~s^{-1}}$; (7) 0.5-8 keV unabsorbed luminosity, in units of $10^{37}{\rm~erg~s^{-1}}$; 
(8) Source variability, defined in text;  (9) Hardness ratio between the 0.5-2 and 2-8 keV bands; (10) Effective radius, in units of pc; (11) Absolute V-band magnitude; 
(12) Metallicity relative to Solar. NGC\,1399-EO12 \citep{Mieske2008}, M104-EO01 \citep{Hau2009}, M59-UCD3 \citep{Sandoval2015}, M60-UCD1 \citep{Strader2013}, NGC\,5128-EO03 \citep{Chattopadhyay2009} and NGC\,5128-EO05, NGC\,5128-EO01 \citep{Mieske2008} have spectroscopically derived metallicity. We estimate metallicity for the other sources based on their color (when available in the literature), assuming a simple stellar population with age of 10 Gyr and a Kroupa IMF.}
\end{deluxetable}


\begin{thebibliography}{}
\bibitem[Bekki et al.(2001)]{Bekki2001} Bekki, K., Couch, W. J., \& Drinkwater, M. J., 2001, \apj, 522, L105
\bibitem[Blakeslee et al.(2012)]{Blakeslee2012} Blakeslee, J. P., Cho, H., Peng, E. W., Ferrarese, L., et al. 2012, \apj, 746, 88
\bibitem[Blom et al.(2012)]{Blom2012} Blom, C., Spitler, L. R., \& Forbes, D. A., 2012, \mnras, 420, 37
\bibitem[Brodie \& Larsen(2002)]{Blom2002} Brodie, J. P., \& Larsen, S. S., 2002, \apj, 124, 1410
\bibitem[Br\"{u}ns et al.(2011)]{Bruns2011} Br\"{u}ns, R. C., Kroupa, P., Fellhauer, M., Metz, M., \& Assmann, P., 2011, A\&A, 529, 138
\bibitem[Br\"{u}ns \& Kroupa(2012)]{Bruns2012} Br\"{u}ns, R. C., \& Kroupa, P., 2012, \aap, 547, A65 (BK12)
\bibitem[Caso et al.(2013)]{Caso2013} Caso, J. P., Bassino, L. P., Richtler, T., Smith Castelli, A. V., \& Faifer, F. R., 2013, \mnras, 430, 1088
\bibitem[Chattopadhyay et al.(2009)]{Chattopadhyay2009} Chattopadhyay, A. K., Chattopadhyay, T., Davoust, E., et al. 2009, \apj, 705, 1533
\bibitem[Chiboucas et al.(2011)]{Chiboucas2011} Chiboucas, K., Tully, R. B., Marzke, R. O., et al. 2011, \apj, 737, 86
\bibitem[Chilingarian \& Mamon(2008)]{Chiling2008} Chilingarian, I. V., \& Mamon, G. A., 2008, MNRAS, 385, L83
\bibitem[Clark(1975)]{Clark1975} Clark, G. W., 1975, \apj, 199, L143
\bibitem[Da Rochar et al.(2011)]{Da Rocha2011} Da Rocha, C., Mieske, S., Georgiev, I. Y., Hilker, M., et al. 2011, \aap, 525, 86
\bibitem[Drinkwater et al.(2000)]{Drinkwater2000} Drinkwater, M. J., Jones, J. B., Gregg, M. D., \& Phillipps, S., 2000, \pasa, 17, 227
\bibitem[Drinkwater et al.(2003)]{Drinkwater2003} Drinkwater, M. J., Gregg, M. D., Hilker, M., et al. 2003, \nat, 423, 519
\bibitem[Drinkwater et al.(2004)]{Drinkwater2004} Drinkwater, M. J., Gregg, M. D., Couch, W. J., et al. 2004, \pasa, 21, 375
\bibitem[Fabbiano(2006)]{Fabbiano2006} Fabbiano, G., 2006, \araa, 44, 323
\bibitem[Fabian(1975)]{Fabian1975} Fabian, A. C., 1975, \mnras, 173, 161
\bibitem[Fellhauer \& Kroupa(2002)]{Fellhauer2002} Fellhauer, M., \& Kroupa, P., 2002, \mnras, 330, 642
\bibitem[Forbes et al.(2013)]{Forbes2013} Forbes, D.A., Pota, V., Usher, C., et al. 2013, MNRAS, 435, L6
\bibitem[Forbes et al.(2014)]{Forbes2014} Forbes, D. A., Almeida, A., Spitler, L. R., \& Pota, V., 2014, \mnras, 442, 1049
\bibitem[Gilfanov(2004)]{Gilfanov2004} Gilfanov, M., 2004, \mnras, 349, 146
\bibitem[Gregg et al.(2009)]{Gregg2009} Gregg, M. D., Drinkwater, M. J., Evstigneeva, E., et al. 2009, \apj, 137, 498
\bibitem[Harris et al.(2012)]{Harris2012} Harris, G. L. H., G\'{o}mez, M., Harris, W. E., Johnston, K., et al. 2012, \aj, 143, 84
\bibitem[Hasegan et al.(2005)]{Hasegan2005} Hasegan, M., Jordan, A., Cote, P., et al. 2005, \apj, 627, 203
\bibitem[Hau et al.(2009)]{Hau2009} Hau, G. K. T., Spitler, L. R., Forbes, D. A., et al. 2009, \mnras, 394, L97
\bibitem[Hilker et al.(1999)]{Hilker1999} Hilker, M., Infante, L., Vieira, G., Kissler-P. M., \& Richtler, T., 1999, \aaps, 134, 75
\bibitem[Hills(1976)]{Hills1976} Hills, J. G., 1975, \mnras, 175, 1
\bibitem[Janz et al.(2016)]{Janz2016} Janz, J., Norris, M.A., Forbes, D.A., et al. 2016, MNRAS, 456, 617
\bibitem[Kalberla et al.(2005)]{Kalberla2005} Kalberla, P.M., Burton, W.B., Hartmann, D., Arnal, E.M., Bajaja, E., Morras R., P\"{o}ppel, W.G.L. 2005, A\&A, 440, 775
\bibitem[Katz(1975)]{Katz1975} Katz, J. I., 1975, \nat, 253, 698
\bibitem[Kim et al.(2007)]{Kim2007} Kim, M., Kim, D.-W., Wilkes, B. J., et al. \apjs, 169, 401
\bibitem[Li et al.(2010)]{Li2010} Li, Z., Spitler, L. R., Jones, C., et al. \apj, 721, 1368
\bibitem[Liu et al.(2015)]{Liu2015} Liu, C.-Z., Peng, E. W., Toloba, E., et al. 2015, \apj, 812, L2
\bibitem[Liu et al.(2015)]{Liu20151} Liu, C.-Z., Peng, E. W., C\^{o}t\'{e}, P., et al. 2015, \apj, 812, 34
\bibitem[Maccarone et al.(2007)]{Maccarone2007} Maccarone, T., Kundu, A., Zepf, S. E., Rhode, K. L. 2007, Nature, 445, 183
\bibitem[Madrid et al.(2010)]{Madrid2010} Madrid, J. P., Graham, A. W., Harris, W. E., et al. 2010, \apj, 722, 1707
\bibitem[Madrid et al.(2013)]{Madrid2013} Madrid, J. P., \& Donzelli, C. J., 2013, \apj, 770, 158
\bibitem[Mieske et al.(2002)]{Mieske2002} Mieske, S., Hilker, M., \& Infante, L., 2002, \aap, 383, 823
\bibitem[Mieske et al.(2007)]{Mieske2007} Mieske, S., Hilker, M., Jord\'{a}n, A., Infante, L., \& Kissler-P., M., 2007, \aap, 472, 111
\bibitem[Mieske et al.(2008)]{Mieske2008} Mieske, S., Hilker, M., Jord\'{a}n, A., Infante, L., et al. 2008, \aap, 487, 921
\bibitem[Mieske et al.(2012)]{Mieske2012} Mieske, S., Hilker, M., \& Misgeld, I., 2012, \aap, 537, A3
\bibitem[Misgeld et al.(2011)]{Misgeld2011} Misgeld, I., Mieske, S., Hilker, M., Richtler, T., et al. 2011, \aap, 531, A4
\bibitem[Norris et al.(2011)]{Norris2011} Norris, M. A., \& Kannappan, S. J., 2011, \mnras, 414, 739
\bibitem[Norris et al.(2015)]{Norris2015} Norris, M. A., Escudero, C. G., Faifer, F. R., et al. 2015, \mnras, 451, 3615
\bibitem[Pandya et al.(2016)]{Pandya2016} Pandya, V., Mulchaey, J., Greene, J.E. 2016, arXiv:1601.01690
\bibitem[Park et al.(2006)]{Park2006} Park, T.-Y.; Kashyap, V. L.; Siemiginowska, A., et al. 2006, \apj, 652, 601
\bibitem[Paudel et al.(2010)]{Paudel2010} Paudel, S., Lisker, T., \& Janz, J., 2010, \apj, 724, L64
\bibitem[Penny et al.(2014)]{Penny2014} Penny, S. J., Forbes, D. A., Strader, J., et al. 2014, \mnras, 439, 3808
\bibitem[Pfeffer \& Baumgardt(2013)]{Pfeffer2013} Pfeffer, J., \& Baumgardt, H., 2013, \mnras, 433, 1997
\bibitem[Phillipps et al.(2001)]{Phillipps2001} Phillipps, S., Drinkwater, M. J., Gregg, M. D., \& Jones, J. B., 2001, \apj, 560, 201
\bibitem[Sandoval et al.(2015)]{Sandoval2015} Sandoval, M. A.; Vo, R. P.; Romanowsky, A. J., et al. 2015, \apj, 808L, 32
\bibitem[Seth et al.(2014)]{Seth2014} Seth, A. C., van den Bosch, R., Mieske, S., Baumgardt, H., Brok, M. D., et al. 2014, \nat, 513, 398
\bibitem[Spitler et al.(2006)]{Spitler2006} Spitler, L. R., Larsen, S. S., Strader, J., Brodie, J. P., et al. 2006, \aj, 132, 1593
\bibitem[Strader et al.(2012)]{Strader2012} Strader, J., Chomiuk, L., Maccarone, T. J., et al. 2012, \nat, 490, 71
\bibitem[Strader et al.(2013)]{Strader2013} Strader, J., Seth, A. C., Forbes, D. A., et al. 2013, \apj, 775, L6
\bibitem[Sutantyo(1975)]{Sutantyo1975} Sutantyo, W., 1975, \aap, 44, 227
\bibitem[Wang(2004)]{Wang2004} Wang, Q.D., 2004, \apj, 612, 159
\bibitem[Zhang et al.(2015)]{Zhang2015} Zhang, H.-X., Peng, E. W., C\^{o}t\'{e}, P., Liu, C.-Z., et al. 2015, \apj, 802, 30

\end{thebibliography}
\end{document}